\documentclass[aps,prd,preprint,nofootinbib]{revtex4}

\usepackage{graphicx}
\usepackage{psfrag}
\usepackage{epsfig}

\begin{document}

\title{Constraints on the Leading-Twist Pion Distribution Amplitude from A QCD
Light-Cone Sum Rule with Chiral Current}
\author{Xing-Gang Wu\footnote{email: wuxg@cqu.edu.cn}}
\address{Department of Physics, Chongqing University, Chongqing 400044,
P.R. China}

\begin{abstract}

We present an improved analysis of the constraints on the first two
Gegenbauer moments, $a^\pi_2$ and $a^\pi_4$, of the pion's
leading-twist distribution amplitude from a QCD light-cone sum rule
analysis of $B\to\pi$ weak transition form factor $f_{+}(q^2)$.
Proper chiral current is adopted in QCD light-cone sum rule so as to
eliminate the most uncertain twist-3 contributions to $f_{+}(q^2)$,
and then we concentrate our attention on the properties of the
leading-twist pion DA. A nearly model-independent $f_+(q^2)$ as
shown in Ref.\cite{pball0} that is based on the spectrum of $B\to\pi
l\nu$ decays from BaBar, together with their uncertainties, are
adopted as the standard shape for $f_+(q^2)$ to do our discussion.
From a minimum $\chi^2$-fit and by taking the theoretical
uncertainties into account, we obtain
$a^\pi_2(1GeV)=0.17^{+0.15}_{-0.17}$ and
$a^\pi_4(1GeV)=-0.06^{+0.20}_{-0.22}$ at the $1\sigma$ confidence
level for $m^*_b\in[4.7,4,8]\; GeV$. \\


\end{abstract}

\maketitle

The pion distribution amplitude (DA) that enters into the exclusive
processes via the factorization theorem at high momentum transfer is
an important factor in perturbative QCD. The leading twist pion DA
is usually expressed in terms of its conformal expansion
\begin{equation}\label{eq:conformal}
\phi_\pi(x,\mu) = 6x(1-x) \left( 1 + \sum_{n=1}^\infty
a_{2n}^\pi(\mu) C^{3/2}_{2n}(2x-1)\right),
\end{equation}
where $x\in[0,1]$ is the momentum fraction of the quark in the pion.
$C^{3/2}_{2n}(2x-1)$ are Gegenbauer polynomials and
$a^\pi_{2n}(\mu)$, the so-called Gegenbauer moments, are hadronic
parameters that depend on the factorization scale $\mu$. Many works
are presented to provide precise values for these Gegenbauer
moments, but till now, whether the pion's leading twist DA is
asymptotic like \cite{brodsky} or CZ-like \cite{cz} is still an open
question, a simple review of this issue can be found in
Ref.\cite{vlz}. Calculations of the second Gegenbauer moment
$a^\pi_{2}$ of pion DA have attracted quite a bit of attentions and
has been discussed through different approaches, a summary of them
can be found in Ref.\cite{lenz} and references therein. Recently,
through a comprehensive analysis of the pion-photon transition from
factor $F_{\pi\gamma}$ involving the transverse momentum corrections
with the CLEO experimental data \cite{cleo}, in which the the
contributions beyond the leading Fock state have been taken into
consideration, Ref.\cite{huangwu1} shows that
$a_2(4GeV^2)=0.002^{+0.063}_{-0.054}$ and
$a_4(4GeV^2)=-0.022_{-0.012}^{+0.026}$ that are closed to the
asymptotic-like behavior of the pion DA.

The process $B\to\pi\ell\nu$ provides a good platform for studying
the pionic distributions. The QCD light-cone sum rule (LCSR)
provides a useful way to study its key factor, i.e. the $B\to\pi$
transition form factor, in the large and intermediate energy regions
($q^2\lesssim 16GeV^2$)\footnote{A consistent analysis of the
$B\to\pi$ form factor in its whole physical region by analyzing the
perturbative QCD, LCSR and Lattice QCD results can be found in
Ref.\cite{huangwu2}.}. By taking the conventional correlation
function for the $B\to\pi$ transition form factors
\cite{pball2,pball3}, it is found that the main uncertainties in
estimation of the $B\to\pi$ transition form factors come from the
different twist structures of the pion wave functions, e.g. the
twist-2 and twist-3 contributions have the same importance. So to
extract more reliable information of the leading-twist DA, one needs
a better understanding of the twist-3 contribution. A comprehensive
analysis calculated from QCD sum rules on the light-cone to
$O(\alpha_s)$ accuracy for twist-2 and the dominant twist-3
contributions has been presented in Refs.\cite{pball2,pball3}, and
it is found at the $1\sigma$ confidence level that \cite{pball1}
\begin{equation}\label{pball}
a^\pi_2(1GeV)=0.19\pm0.19 , \quad a^\pi_4(1GeV)\geq-0.7 \ .
\end{equation}
On the other hand, it has been found that by choosing proper chiral
currents in the LCSR approach, the contributions from the most
uncertain twist-3 structures to the form factor can be directly
eliminated \cite{huangbpi1,huangbpi2}. In Ref.\cite{wuhuang} we have
shown that these two treatments to deal with the twist-3
contributions of the $B\to\pi$ or $B\to K$ form factors are
equivalent to each other. Since the pollution from the twist-3
structures are eliminated and the even higher twist structures
provide small contributions (less than $5\%$), so the LCSR with
chiral current may derive more precise information on the leading
twist-2 DA. This is the purpose of the present letter. Furthermore,
our present analysis shall also provide a meaningful cross check of
$a^\pi_2$ and $a^\pi_4$ derived in Ref.\cite{pball1} through the
conventional LCSR calculation.

The hadronic matrix element relevant for $B\to\pi\ell\nu$ is given
by
\begin{equation}
\langle \pi(p_\pi)| \bar u \gamma_\mu b | B(p_B)\rangle = \left(
p_{B\mu} + p_{\pi\mu} - \frac{m_B^2-m_\pi^2}{q^2}q_\mu\right)
f_{+}(q^2) + \frac{m_B^2-m_\pi^2}{q^2}\, q_{\mu} f_0(q^2),
\end{equation}
where the form factors $f_{+,0}$ depend on $q^2\equiv(p_B-p_\pi)^2$,
the invariant mass of the lepton-pair, with $0 \leq q^2\leq
(m_B-m_\pi)^2\simeq 26.4 $GeV$^2$. Only $f_+(q^2)$ is needed for
calculating the spectrum, i.e.
\begin{equation}
\frac{d\Gamma}{dq^2}\,(B^0\to \pi^- \ell^+ \nu_\ell) = \frac{G_F^2
  |V_{ub}|^2}{ 192 \pi^3 m_B^3}\,\lambda^{3/2}(q^2) |f_+(q^2)|^2
\end{equation}
for massless leptons, where $\lambda(q^2) = (m_B^2+m_\pi^2-q^2)^2 -
4 m_B^2 m_\pi^2$ is the usual phase-space factor. By taking the LCSR
with chiral current, it is found that the main theoretical
uncertainty comes from the pion's leading-twist light-cone DA
$\phi_\pi$, and other smaller uncertainty sources include the $b$
quark mass, the quark condensate, sum rule specific parameters
(Borel parameter and continuum threshold) and etc. Numerically, it
can be found that the $q^2$-dependence of the form factor
$f_{+}(q^2)$ is mostly sensitive to $a_2^\pi$ and only to a lesser
extent to higher Gegenbauer-moments. We hence decide to use the
$\phi_\pi$ proposed in Eq.(\ref{eq:conformal}), which we truncate
after the contribution in $a_4^\pi$.

\begin{figure}
\centering
\includegraphics[width=0.45\textwidth]{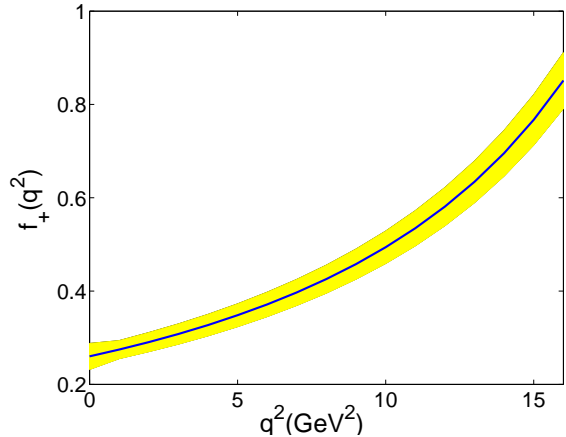}
\caption{Best fits for $f_{+}(q^2)$ that are derived from the
fitting of the BaBar experimental data \cite{babar}, where the solid
line is for BK parameterization \cite{pball0}. The shaded band shows
the total uncertainties that include the errors of the five
parameterizations.} \label{ballfit}
\end{figure}

A nearly model-independent analysis for $f_{+}(q^2)$ based on the
experimental data has been given in Ref.\cite{pball0}, in which the
value of $V_{ub}$ from the UTfit Collaboration \cite{utfit} and the
CKMfitter Collaboration \cite{ckmfit}, e.g.
$|V_{ub}|=(3.50\pm0.18)\times 10^{-3}$, and the spectrum of $B\to\pi
l\nu$ decays from BaBar \cite{babar} have been adopted. The best
fits obtained by using five parameterizations of $f_{+}(q^2)$, i.e.
Becirevic/Kaidalov (BK)\cite{bk}, Ball/Zwicky(BZ)\cite{bz},
Boyd/Grinstein/Lebed(BGL)\cite{bgl} with two choices for its free
parameter $q^2_0$ (called as BGLa or BGLb parameterization
respectively), and the Omnes representation of Ref.\cite{afhnv}
(AFHNV), are very close to each other in the low and intermediate
energy regions (all best-fit form factors agree within $2\%$
\cite{pball0}) and noticeable differences occur only for large $q^2$
region. Since the QCD LCSR are reliable only in low and intermediate
energy regions that is less than $16GeV^2$, so we shall adopt the
fitted results of these five parameterizations with their possible
errors within the region of $q^2<16GeV^2$ as the standard shape for
$f_+(q^2)$ to do our following discussion. We shall not extrapolate
our LCSR result to even higher energy regions in order to minimize
any uncertainty from extrapolating in $q^2$. More explicitly, the
best fits obtained by using those five parameterizations of
$f_{+}(q^2)$ shown in Fig.(\ref{ballfit}), together with the
additional $\pm 3\%$ error from the total branching ratio of
$B\to\pi$, shall be used as the experimentally determined shape of
the form factor. Further more, since the center value of the above
listed five parameterizations are very close to each other in the
region of $q^2<16GeV^2$, so we take the simpler BK-parameterization
to be the center value of $f_+(q^2)$ as shown by the solid line of
Fig.(\ref{ballfit}), i.e.
\begin{equation}\label{BK}
\left.f_+(q^2)\right|_{{\rm BK}} =\frac{f_+(0)}{
(1-q^2/m_{B^*}^2)(1-\alpha\,q^2/m_{B^*}^2)}
\end{equation}
with $m_{B^*}=5.325GeV$, $f_+(0)=0.26$ and $\alpha=0.53$
\cite{pball0}. Then by fitting our LCSR result with chiral current
to the experimentally determined shape for $f_{+}(q^2)$, we can
determine the possible regions of $a^\pi_2$ and $a^\pi_4$.

\begin{table} \centering
\begin{tabular}{|c||c|c|c||c|c|c|}
\hline ~~~ & \multicolumn{3}{|c||}{LO result} &  \multicolumn{3}{|c|}{NLO result} \\
\hline ~~~ - ~~~ & ~~~$s_0$~~~ & ~~~$M^2$~~~ & ~~~ $f_B$ ~~~&
~~~$s_0$~~~ & ~~~$M^2$~~~ & ~~~ $f_B$ ~~~\\
\hline\hline
~~$m^*_b=4.7$~~ & 33.5 & 2.80 & 0.165 & 33.5 & 2.80 & 0.219 \\
\hline $m^*_b=4.8$ & 33.2 & 2.39 & 0.131 & 33.2 & 2.31 & 0.174\\
\hline
\end{tabular}
\caption{Parameters for $f_B$, where $m^*_b$ and $f_B$ are given in
$GeV$, $s_0$ and $M^2$ in $GeV^2$. The first direct measurement of
$f_B$ by Belle experiment shows $f_B=229^{+36}_{-31}({\rm stat.})
^{+34}_{-37}({\rm syst.})$ MeV \cite{belle}. }\label{tabfb}
\end{table}

Before a comparison of our LCSR result with the fitted shape for
$f_{+}(q^2)$, we make some comments on the treatment of $f_B$. To be
consistent, $f_B$ should be varied accordingly and be determined by
using the two-point sum rule with the chiral currents. The sum rule
for $f_B$ up to NLO can be obtained from Ref.\cite{sumrulefb}
through a proper combination of the scalar and pseudo-scalar results
shown there, which can be schematically written as
\begin{equation} \label{sumfb}
f_B^2 M_B^2 e^{-M^2_B/M^2}=\int_{m_b^2}^{s_0}\rho^{tot}(s)
e^{-s/M^2}ds ,
\end{equation}
where the spectral density $\rho^{tot}(s)$ can be read from
Ref.\cite{sumrulefb}. The Borel parameter $M^2$ and the continuum
threshold $s_0$ are determined such that the resulting form factor
does not depend too much on the precise values of these parameters;
in addition, 1) the continuum contribution, that is the part of the
dispersive integral from $s_0$ to $\infty$, should not be too large,
e.g. less than $20\%$ of the total dispersive integral; 2) the
contributions from the dimension-six condensate terms shall not
exceed $15\%$ for $f_B$; 3) the derivative of the logarithm of
Eq.(\ref{sumfb}) with respect to $1/M^2$ gives the B-meson mass
$M_B$ \cite{bz},
\begin{displaymath}
M_B^2=\int_{m_b^2}^{s_0}\rho^{tot}(s) e^{-s/M^2} s ds {\Bigg /}
\int_{m_b^2}^{s_0}\rho^{tot}(s) e^{-s/M^2}ds ,
\end{displaymath}
and we require its value to be full-filled with high accuracy $\sim
0.1\%$. These criteria define a set of parameters for each value of
the effective $b$-quark $m^*_b$ and some typical values are listed
in Tab.\ref{tabfb}, where $f_B$ is taken as the extremum within
reasonable region of $(M^2,s_0)$.

The LCSR with chiral current for $f_{+}(q^2)$ including twist-2
contributions to ${\cal O}(\alpha_s)$ accuracy and twist-4
contributions at tree-level can be found in
Refs.\cite{wanghuang,wuhuang,khod}. The interesting reader may turn
to these references for more detailed technology, especially the
$B\to\pi$ form factor can be directly obtained from
Ref.\cite{wuhuang} by properly ignoring the $SU_f(3)$-breaking
effect in the $B\to K$ form factor. As a comparison, we obtain
values for $f_{+}(q^2)$ in dependence of $a^\pi_2$, $a^\pi_4$ and
$m^*_b$ using the same criteria as suggested in Ref.\cite{pball1}
for the evolution of the LCSRs. Further more, for each value of
$f_+(q^2)$ we calculate the theoretical uncertainty by varying: 1)
the Borel parameter $M^2$ in the LCSR for $f_+(q^2)$ within the
region of $[10, 18]\; GeV^2$; 2) $s_0$ by $\pm 1\,$GeV$^2$; 3) the
central value 20\% of the continuum contribution between 15\% and
25\%. The above ranges of sum rule parameters are rather
conservative and account for the ``systematic'' uncertainty of QCD
sum rule calculations.

Next, we require the LCSR result to be compatible with the BaBar
experimental data, i.e. the LCSR result of $f_+(q^2)$ should be
within the shaded band of Fig.(\ref{ballfit}) with $q^2\leq 16
GeV^2$, then we can derive the reasonable ranges for $a^\pi_2$,
$a^\pi_4$ and $m_b^*$. Further more, we adopt $a^\pi_2(1GeV)\geq 0$
that is favored in the literature \cite{moments}, and $m_b^*\in
[4.7,4.8] GeV$ \cite{pball1} \footnote{A review of Ref.\cite{mbmass}
shows that $m_b^*\simeq 4.8 \pm 0.1GeV$, which can also be adopted
by allowing the discrepancy between the LCSR and the PQCD
calculation in the low energy region to be less than $15\%$
\cite{wuhuang}.} to do our discussions.

\begin{figure}
\centering
\includegraphics[width=0.45\textwidth]{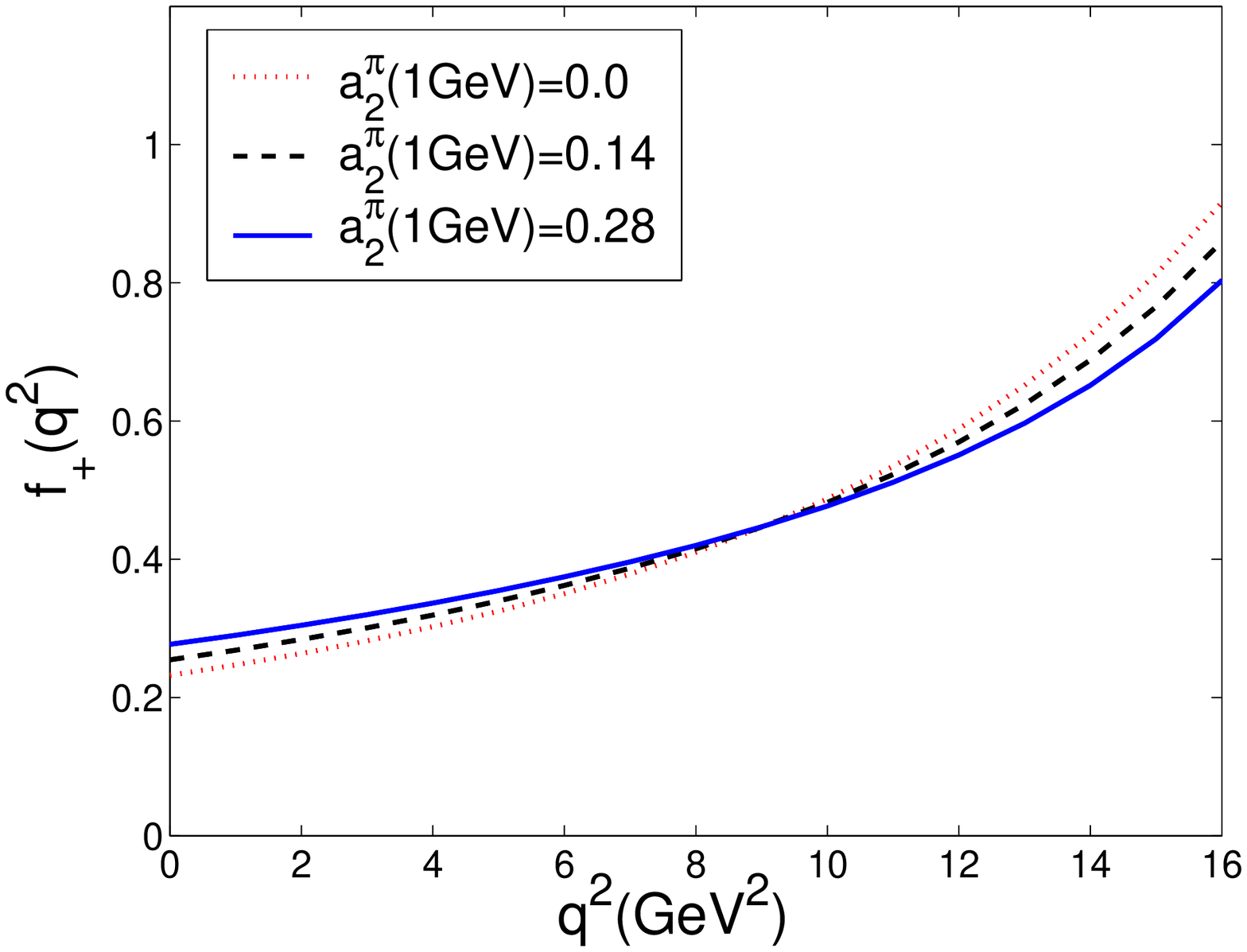}%
\hspace{0.5cm}
\includegraphics[width=0.45\textwidth]{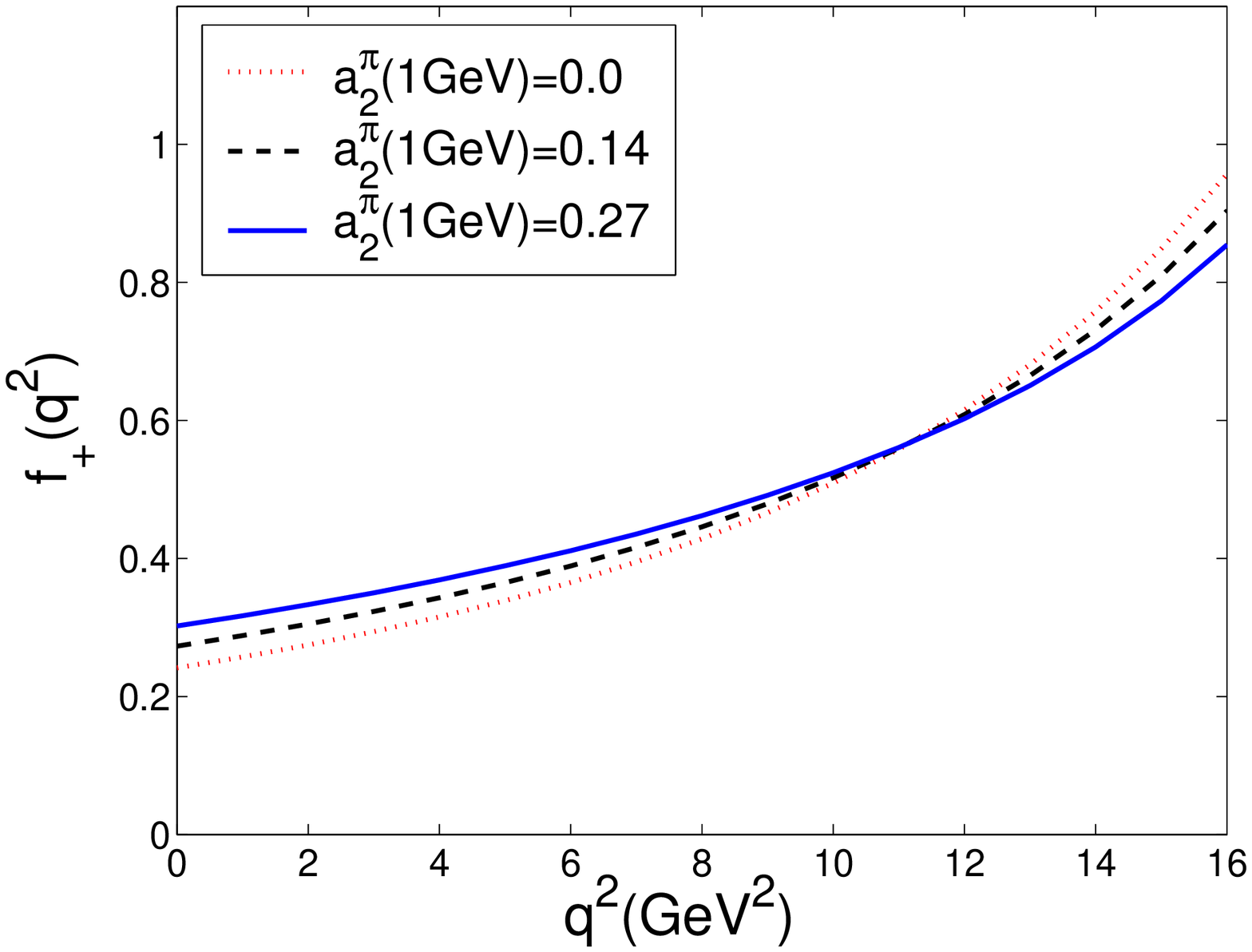}%
\caption{$f_{+}(q^2)$ of $a^\pi_2$ for the case of $a^\pi_4=0$,
$m_b^*=4.7GeV$ (Left) and $m_b^*=4.8GeV$ (Right). $f_{+}(q^2)$ {\it
increases} with the increment of $a^\pi_2$ in the lower energy
region but {\it decreases} with the increment of $a^\pi_2$ in the
higher energy region.} \label{a2pi}
\end{figure}

\begin{figure}
\centering
\includegraphics[width=0.45\textwidth]{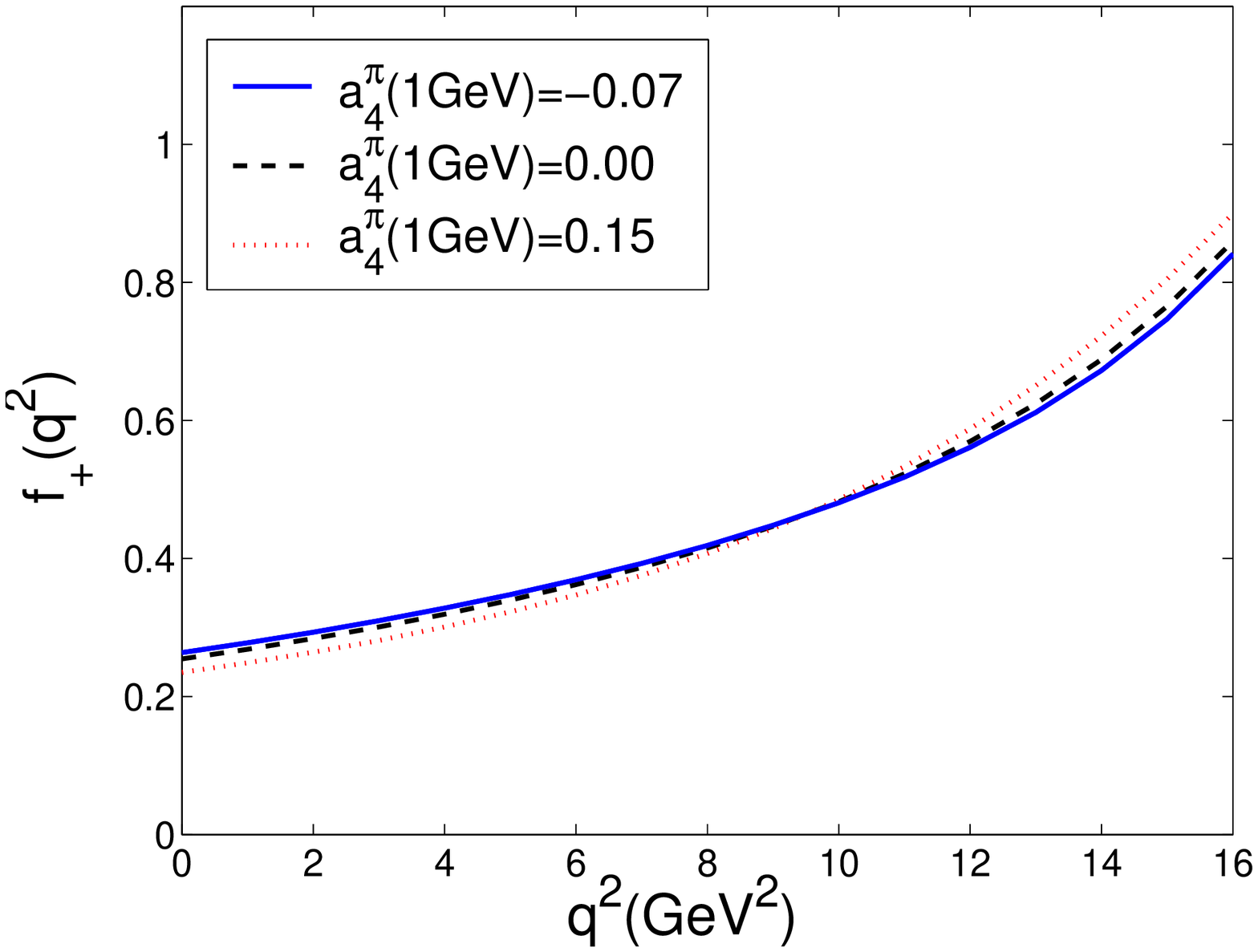}%
\hspace{0.5cm}
\includegraphics[width=0.45\textwidth]{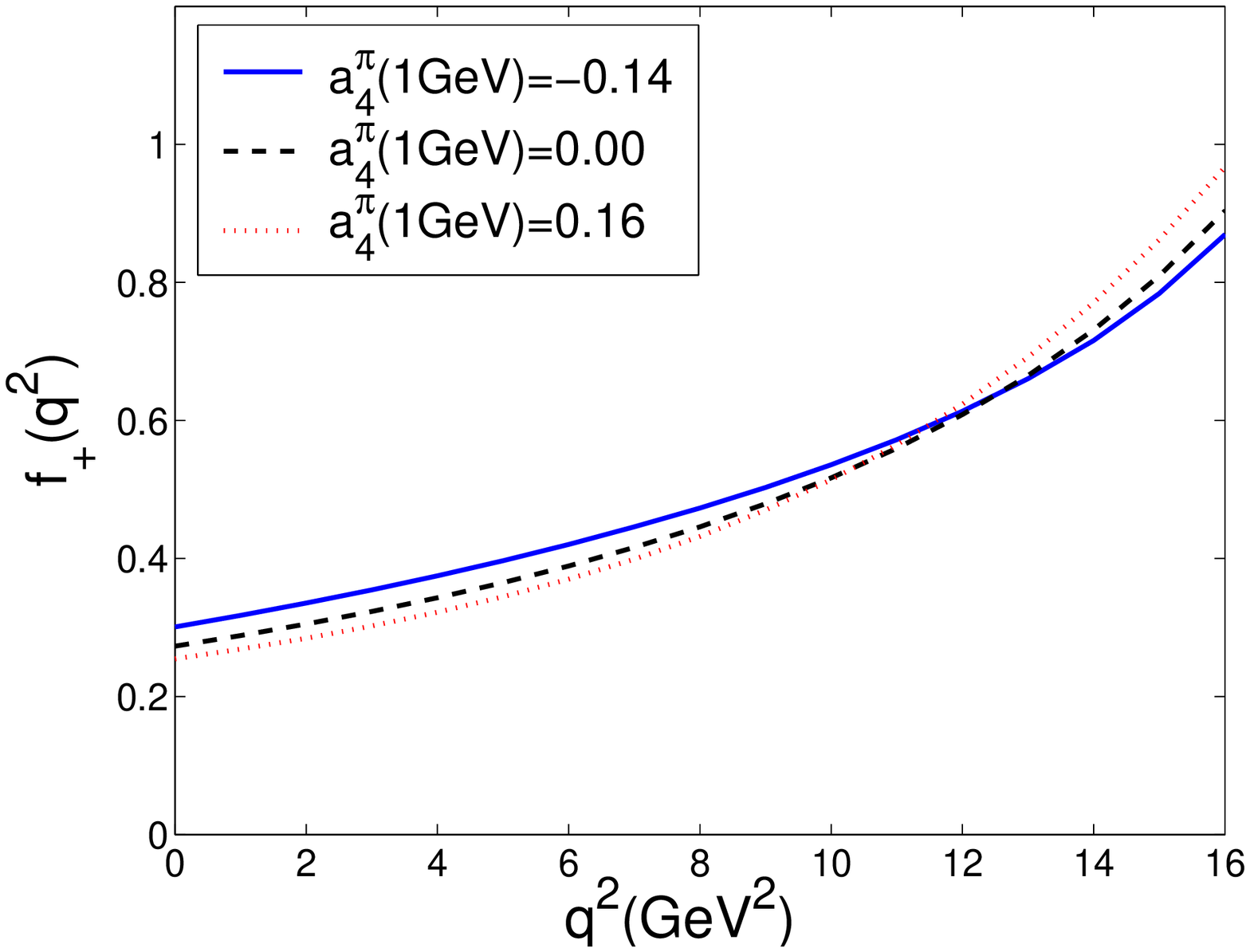}%
\caption{$f_{+}(q^2)$ of $a^\pi_4$ for the case of $a^\pi_2=0.14$,
$m_b^*=4.7GeV$ (Left) and $m_b^*=4.8GeV$ (Right). $f_{+}(q^2)$ {\it
decreases} with the increment of $a^\pi_4$ in the lower energy
region but {\it increases} with the increment of $a^\pi_4$ in the
higher energy region. } \label{a4pi}
\end{figure}

Before deriving the possible ranges of these parameters, we show the
properties of $f_+(q^2)$ versus $a^\pi_2$ and $a^\pi_4$ respectively
in Figs.(\ref{a2pi}, \ref{a4pi}) by varying $a^\pi_2$ and $a^\pi_4$
independently. From Figs.(\ref{a2pi}, \ref{a4pi}), it can be found
that $f_{+}(q^2)$ {\it increases} ({\it decreases}) with the
increment of $a^\pi_2$ ($a^\pi_4$) in the lower energy region but
{\it decreases} ({\it increases}) with the increment of $a^\pi_2$
($a^\pi_4$) in the higher energy region, so possible range of
$a^\pi_2$ or $a^\pi_4$ can indeed be derived by demanding $f_+(q^2)$
within the fitted band of Fig.(\ref{ballfit}). Furthermore, since
$f_{+}(q^2)$ increases with the increment of $m^*_b$, the value of
$m^*_b$ should not be too large.

\begin{figure}
\centering
\includegraphics[width=0.45\textwidth]{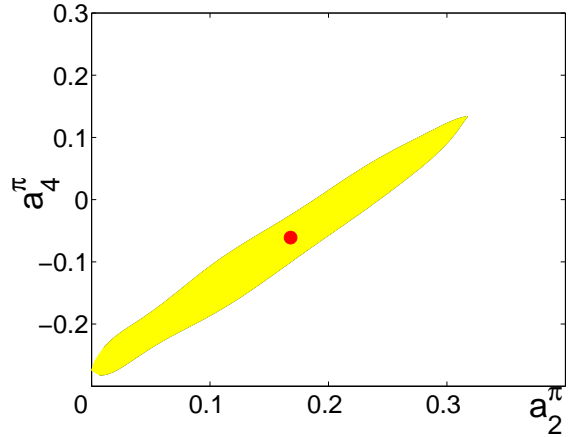}
\caption{Allowed values of $a_{2,4}^\pi(1GeV)$ for $m_b \in [4.7,
4.8] {\rm GeV}$ with $\chi^2 \leq \chi^2_{\alpha=68.27\%}(3)=1.87$
for 3 d.o.f. at the $1\sigma$ C.L..} \label{chi}
\end{figure}

Now the task is to compare the form factor prediction to data and to
determine best-fit values of $a^\pi_2$ and $a^\pi_4$, using the
experimentally determined shape for $f_{+}(q^2)$ as shown in
Fig.(\ref{ballfit}). More explicitly, we take one hundred
$f_{+}(q^2)$ points uniformly within the range of $q^2\in [0, 16]
GeV^2$ respectively, together with their corresponding errors that
can be derived from Ref.\cite{pball0} to do the calculation. The
resulting constraints are shown in Fig.(\ref{chi}). The minimum
$\chi^2$ for $(a_2^\pi,a_4^\pi)$ is reached for $a_2^\pi = 0.17$ and
$a_4^\pi =-0.06$, i.e. near the center of the parameter space. The
biggest counter include all $(a_2^\pi,a_4^\pi)$ for which the fit of
the corresponding form factor to the data yields $\chi^2\leq
\chi^2_{\alpha=68.27\%}(3)=1.87$ for 3 degree of freedom (d.o.f.) at
the $68.27\%$ ($1\sigma$) confidence level (C.L.), where
$\chi^2_{\alpha}(n)$ is derived from
$\int^{\chi^2_{\alpha}(n)}_0f(y;n)dy=\alpha$ with
$f(y;n)=\frac{1}{\Gamma(n/2)2^{n/2}}y^{n/2-1}e^{-y/2}$. One can
immediately read off the following constraints of $a^\pi_{2,4}$ at
the $1\sigma$ confidence level
\begin{equation}\label{wu}
a^\pi_2(1GeV)=0.17^{+0.15}_{-0.17},\qquad
a^\pi_4(1GeV)=-0.06^{+0.20}_{-0.22} \;.
\end{equation}
This result is consistent with that of Ref.\cite{pball1}, i.e.
Eq.(\ref{pball}), but with less uncertainty, which shows that the
two independent treatments of the pionic twist-3 contributions are
consistent with each other. This also inversely implies that by
properly taking the parameter values, the LCSR prediction can be
compatible with the experimentally determined shape of the form
factor.

To summarize: we have presented an improved analysis on the pionic
leading twist DA posed by the recently derived fitted shape for the
$B\to\pi$ form factor $f_{+}(q^2)$ from the BaBar experimental data
and the value of $V_{ub}$ from the UTfit and CKMfitter
Collaborations. It is found that the present LCSR with chiral
current are consistent with the conventional LCSR result
\cite{pball1}, and we obtain $a^\pi_2(1GeV)=0.17^{+0.15}_{-0.17}$
and $a^\pi_4(1GeV)=-0.06^{+0.20}_{-0.22}$ at the $1\sigma$
confidence level for $m^*_b\in[4.7,4,8]\; GeV$. Since the twist-3
contribution is eliminated, the present LCSR result is less
uncertain than that of the conventional LCSR analysis. The LCSR sum
rule with chiral current provides a useful way to simply the
conventional LCSR calculation, so it can be applied for other useful
processes, a review of it shall be presented elsewhere
\cite{chiralsr}.

\hspace{2cm}

\begin{center}

{\bf Acknowledgements} \\

\end{center}

This work was supported in part by the grant from Chongqing
University, and by the grant from the Chinese Academy of Engineering
Physics under the grant numbers: 2008T0401 and 2008T0402. \\

\end{document}